\documentclass{sf2a-conf2016}
\usepackage{graphicx}
\usepackage[breaklinks=true]{hyperref}
\hypersetup{
    bookmarksopen=false,
    bookmarksnumbered=false,
    unicode=true,           % non-Latin characters in Acrobat’s bookmarks
    pdftoolbar=true,        % show Acrobat’s toolbar?
    pdfmenubar=true,        % show Acrobat’s menu?
    pdffitwindow=false,     % window fit to page when opened
    pdfstartview={FitH},    % fits the width of the page to the window
    pdfnewwindow=true,      % links in new window
    colorlinks=true,       % false: boxed links; true: colored links
    linkcolor=red,          % color of internal links
    citecolor=magenta,        % color of links to bibliography
    filecolor=green,      % color of file links
    urlcolor=blue           % color of external links
}
\usepackage[]{natbib}  
\usepackage{epstopdf}

\def\BibTeX{{\rm B\kern-.05em{\sc i\kern-.025em b}\kern-.08em
    T\kern-.1667em\lower.7ex\hbox{E}\kern-.125emX}}
\bibpunct{(}{)}{;}{a}{}{,}  %%%%%%%%%%%%%  A&A bibliography style
\begin{document}

\TitreGlobal{SF2A 2016}

%%-----------------------------------------------------------------
%%      the top matter
%%

\title{Re-grouping stars based on the chemical tagging technique: A case study of M67 and IC4651}

\runningtitle{The chemical tagging technique}

\author{S. Blanco-Cuaresma}\address{Observatoire de Gen\`eve, Universit\'e de Gen\`eve, CH-1290 Versoix, Switzerland}

\author{C. Soubiran}\address{Laboratoire d'astrophysique de Bordeaux, Univ. Bordeaux, CNRS, 33615 Pessac, France}

%% IF Author3 has the same affiliation than Author1:
%\author{C.\,E. Author3$^1$}

%% IF Author3 has its own affiliation:
%\author{C.\,E. Author3}\address{Dept. of Chess, University of Games, 35101 Las Vegas, Monaco} 

%% IF Author3 has two affiliations, the one of Author1 and a second one:
%\author{C.\,E. Author3$^{1,}$}\address{Dept. of Chess, University of Games, 35101 Las Vegas, Monaco} 

%% Keep this line, even if the page will be settled afterwards.
\setcounter{page}{237}

%%-----------------------------------------------------------------

\maketitle

%%-----------------------------------------------------------------
%%        The abstract
%% 
%%  Warning!  within the abstract:
%%  - do not use macros. 
%%  - do not use commands like: \cite, \citet, \citep ... etc.

\begin{abstract}
The chemical tagging technique proposed by \cite{2002ARA&A..40..487F} is based on the idea that stars formed from the same molecular cloud should share the same chemical signature. Thus, using only the chemical composition of stars we should be able to re-group the ones that once belonged to the same stellar aggregate. In \cite{2015A&A...577A..47B}, we tested the technique on open cluster stars using iSpec \citep{2014A&A...569A.111B}, we demonstrated their chemical homogeneity but we found that the 14 studied elements lead to chemical signatures too similar to reliably distinguish stars from different clusters. This represents a challenge to the technique and a new question was open: Could the inclusion of other elements help to better distinguish stars from different aggregates? With an updated and improved version of iSpec, we derived abundances for 28 elements using spectra from HARPS, UVES and NARVAL archives for the open clusters M67 and IC4651, and we found that the chemical signatures of both clusters are very similar.
\end{abstract}

%% Insert the keywords (to appear in the ADS indexing)
%% Keywords must be separated by a comma
\begin{keywords}
stars, chemical abundances, metallicity, chemical tagging
\end{keywords}

%%-----------------------------------------------------------------

\section{Introduction}
%%---------------------
  %Enter here the text of your introduction.

The chemical composition of a star provides an invaluable source of information about its history, the stellar aggregate were it was born (in some cases, still gravitationally bounded), the molecular cloud from which it was formed and, finally, the characteristics of that region and time of the Galaxy. It is accepted that most of the stars are born in groups and, if we assume that the original giant molecular cloud was homogeneous and well-mixed, then we can expect that the stars born together share a common chemical fingerprint that may be different from other stellar aggregates (born in different places and times). The chemical tagging technique \citep{2002ARA&A..40..487F} consists in identifying stars that were born together by only looking into their chemical abundances, thus, re-construct the history of our Galaxy.

In \cite{2015A&A...577A..47B}, we designed and executed an experiment using open clusters (most of them with solar metallicities) to test the limits of the chemical tagging technique. We compiled a large dataset of high-resolution stellar spectra from stars in clusters, then we treat each of them as individual isolated stars, we homogeneously derived the chemical abundances for 14 elements and we tried to re-group the stars based only on their chemical information. We found that, given the level of precision that we obtained because of to the spectra quality and the limits of the methods, the differences between different open clusters for the selected elements were not significant enough to correctly disentangle their stars.

In this study, we concentrated our efforts in only two clusters (M67 and IC4651) and we explore the possibility of using more elements to overcome the problems we found in \cite{2015A&A...577A..47B}. Additionally, we developed a new spectroscopic pipeline that takes advantage of the latest improvements implemented in iSpec\footnote{\href{http://www.blancocuaresma.com/s/}{http://www.blancocuaresma.com/s/}} \citep{2014A&A...569A.111B}.

\section{Open clusters}\label{blancocuaresma:sectOpenClusters}
%%%-------------------------

M67 and IC4651 are open clusters with a chemical composition similar to the Sun. They are located in the galactic anti-center (M67: $l=215.70^{\circ}$, $b=31.90 ^{\circ}$; IC4651: $l=340.09^{\circ}$, $b=-7.91^{\circ}$) with a distance from the Sun of 790 pc and 890 pc, respectively \citep{2002A&A...389..871D, 2006MNRAS.371.1641P}.

\section{Data}
%%%-------------------------

Our initial dataset contained 103 spectra for M67 (52 from the UVES archive \citealt{2000SPIE.4008..534D}; 51 from the HARPS archive \citealt{2003Msngr.114...20M}) and 41 for IC4651 (33 from UVES; 8 from HARPS; 1 from NARVAL \citealt{2003EAS.....9..105A}). We measured their radial velocities and compared them to the average velocities reported by HARPS: 33.77 km/s for M67 and -30.36 km/s for IC4651. The criteria to accept a star as member of the cluster was that its velocity should be within the $\pm$2 km/s margin. Only 1 star belonging to IC4651 did not match the criteria and was discarded.

Then, we corrected the radial velocities, co-added spectra for common stars coming from the same instrument and setup, and we selected only those with a signal-to-noise (S/N) higher than 100. We could use spectra with a lower S/N but for this study we wanted to minimize problems originated by low S/N and not linked to real physical features of each star. Once the processing was completed, the dataset was reduced to 22 M67 spectra (all of them observed with UVES) and 16 IC4651 spectra (8 UVES, 7 HARPS and 1 NARVAL) which correspond to 21 M67 stars and 11 IC4651 stars.

\section{Method}
%%%-------------------------

The spectroscopic analysis was done with an automatic pipeline based on iSpec.  In a first step, the atmospheric parameters (i.e. effective temperature, surface gravity and metallicity) were determined for all the co-added spectra by using a selection of absorption lines in the visual range (i.e. 480 to 680 nm).  We used SPECTRUM \citep{1994AJ....107..742G} as radiative transfer code, MARCS\footnote{\url{http://marcs.astro.uu.se/}} \citep{2008A&A...486..951G} as model atmosphere and \cite{2007SSRv..130..105G} as solar abundances. The lines were chosen based on a previous analysis of a solar spectrum (with a resolution of 47\,000) obtained from the Gaia FGK Benchmark Stars library\footnote{\url{http://www.blancocuaresma.com/s/}} \citep{2014A&A...566A..98B} where, starting from an atomic line list extracted from VALD \citep{2011BaltA..20..503K}, we derived abundances ([El/X]) for all the observed absorption lines and we discarded those with a [El/H] greater/smaller than +/-0.05 dex.

In our dataset we found 12 dwarfs, 3 turn-off stars and 8 giants for M67; while for IC4651 we had 3 dwarfs and 8 giants. The criteria to separate dwarfs from giants was based on its surface gravity, stars with gravities greater than 4 dex were considered dwarfs and smaller than 3 dex were classified as giants (in between these limits the star is considered in the turn-off).

Once the atmospheric parameters were fixed, individual chemical abundances were derived using an extensive collection of observed absorption lines which were cross-matched with a VALD atomic line list. For each line, the following abundances were determined:

\begin{enumerate}
    \item Fixed atmospheric parameters (main iteration).
    \item Artificially higher metallicity (+0.10 dex).
    \item New realisation of the spectrum using a poisson distribution and the flux errors (S/N iteration).
    \item New atomic line list where only the line of interest is present (i.e. no blends will be modeled).
\end{enumerate}

We discarded all the lines where the difference with the main iteration is greater than 0.10 dex when using a greater metallicity, greater than 0.10 when checking the S/N or greater than 0.50 dex when synthesized without blends. The accepted lines were combined using the median, a robust statistic to minimize the impact of outliers. Errors were computed using the standard deviation, we avoided the use of more robust statistics (such as the median absolute deviation or MAD) to remain on the conservative side of the estimation.

To achieve the maximum precision, it is common to perform differential analysis. This means that all abundances are computed differentially line-by-line using a star of reference. Typically this is done using a solar spectrum, but this strategy brings up a problem: not all the stars in the dataset have similar atmospheric parameters to those of the Sun. 

Giants and dwarfs are at different stages of their life and there are processes that may have affected their chemical abundances (such as atomic diffusion). Also, assumptions in our analysis (e.g. LTE) may not apply equally (e.g. NLTE effects), or simply continuum normalization is going to be different because the typical spectrum is different (e.g. absorption lines in dwarfs are broader). If we use the Sun as reference, we would derive different chemical signatures for dwarfs and giants, even if they belong to the same cluster. In front of this situation, in \cite{2015A&A...577A..47B} we had to treat giants and dwarfs separately. This is a valid solution but it has the inconvenient of decreasing the statistics.

In this study we tried a different approach, instead of using the Sun as reference for all the stars, we took the giant M67 No164 and the dwarf M67 No1194 as reference stars for the giants and dwarfs in our sample. We assumed that these two stars have the same chemical abundance and that their differences are originated mainly due to assumptions in our models and biases in our analysis, thereby we were able to mix the differential abundances coming from giants and dwarfs.

\section{Results}
%%%-------------------------

\begin{figure}[ht!]
 \centering
 \includegraphics[width=0.8\textwidth,clip]{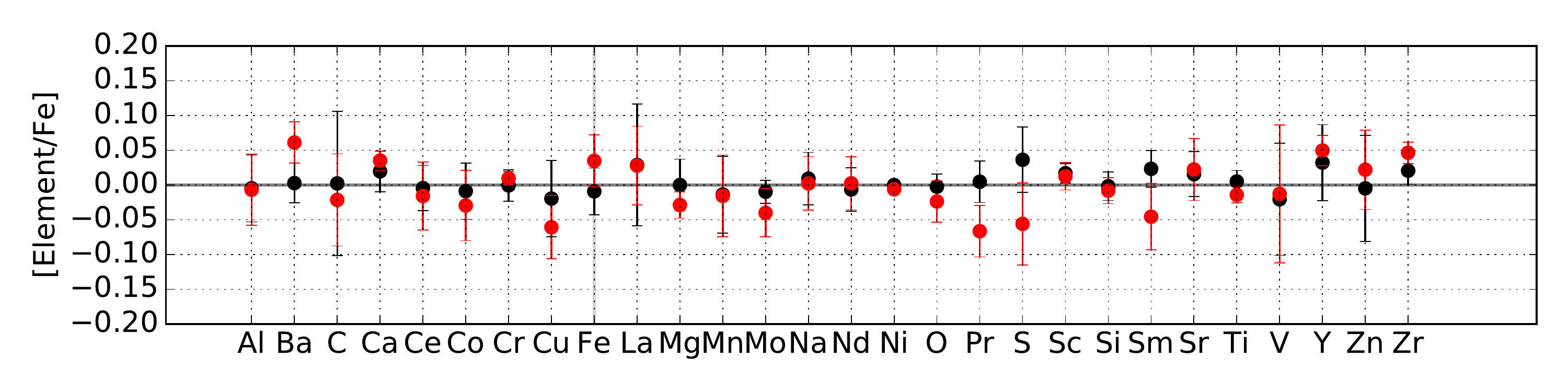}
  \caption{Combined differential chemical abundances with M67 No164 and M67 No1194 as reference for giants and dwarfs, respectively. The chemical signature corresponds to the open clusters M67 (black) and IC4651 (red). All the abundances are respect to iron (i.e. [E/Fe]), except iron which is represented in respect to hydrogen (i.e. [Fe/H]). Error bars correspond to the standard deviation of the abundances.}
  \label{blancocuaresma:fig1}
\end{figure}

In \cite{2015A&A...577A..47B} we were not able to disentangle stars from different clusters by using the chemical signature of 14 elements. The following question was raised: Could a different combination of elements help? In this study we have increased the number of elements to 28 for the open clusters M67 and IC4561 as shown in Fig.~\ref{blancocuaresma:fig1}. Most of the elements have a dispersion smaller than 0.05 dex, demonstrating that the level of precision is high and that the strategy for mixing giants and dwarfs is working properly. Nevertheless, the chemical signatures remain extremely similar between the two clusters. The elements showing the larger differences, such as barium, praseodymium or sulfur, are those which have the largest dispersion, preventing a clear separation of both populations.

\begin{figure}[ht!]
 \centering
 \includegraphics[width=0.8\textwidth,clip]{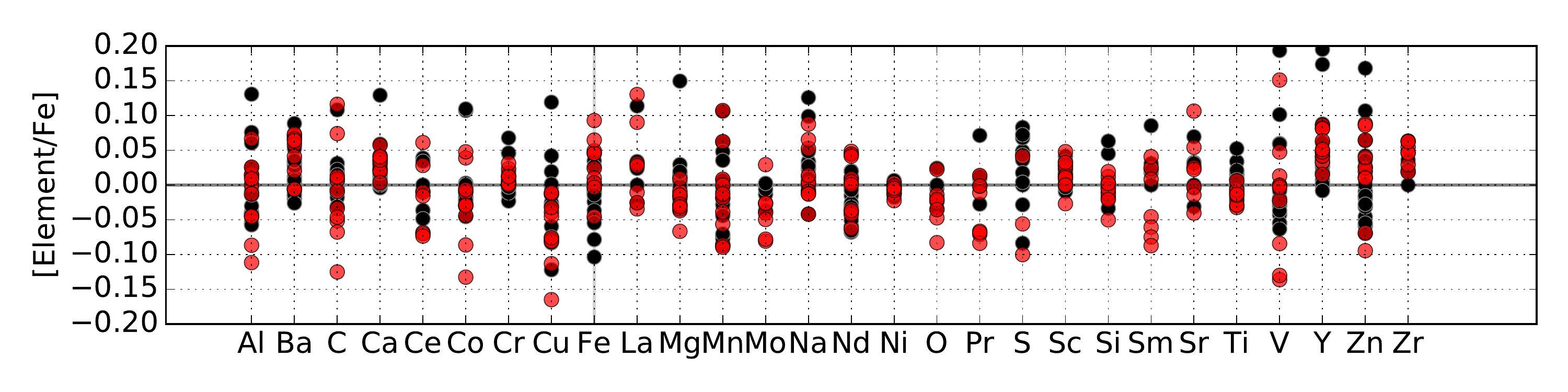}
  \caption{Individual differential chemical abundances with M67 No164 and M67 No1194 as reference for giants and dwarfs, respectively. The abundances correspond to individual stars from the open clusters M67 (black) and IC4651 (transparent red). All the abundances are respect to iron (i.e. [E/Fe]), except iron which is represented in respect to hydrogen (i.e. [Fe/H]).}
  \label{blancocuaresma:fig2}
\end{figure}

Checking directly the individual abundances per star (Fig.~\ref{blancocuaresma:fig2}) shows the great level of overlapping. If we ignore the color coding, it is not feasible to identify what stars belong to what cluster. In this chemical space of 28 elements, M67 and IC4651 are not located in clearly different places.

\section{Conclusions}

We showed how a chemical signature composed of 28 different elements with a general precision better than 0.05 dex does not seem enough to chemically separate stars from the open clusters M67 and IC4651. Is this result still challenging the chemical tagging technique or these two cluster do have a common past? 

Both clusters are located towards the galactic anti-center at a similar distance from the Sun, although they are separated by more than $100^{\circ}$. Additionally, some studies found that M67 is several Gyr\footnote{1 Gyr represents $10^9$ years.} older than IC4651 \citep{2006MNRAS.371.1641P}. Could these clusters be born from the same molecular cloud but at different moments? This would required the cloud to be fragmented in two without triggering star formation and remaining chemically unaltered during a long time. Another possibility would be that it is common to have different molecular clouds with very similar compositions, which could mean that both were enriched in the same measure by different past events. The similarities between these two clusters should be further studied.

% Optional acknowledgements
% -------------------------
\begin{acknowledgements}
This work would not have been possible without the support of Laurent Eyer from the University of Geneva.
\end{acknowledgements}

\bibliographystyle{aa}  % A&A bibliography style file (aa.bst)
\bibliography{ChemicalTagging} % your references in file: Yourfile.bib

\end{document}